\documentclass[12pt]{article}
\usepackage{amsmath,amssymb,amsthm,color}
\usepackage{graphicx}
\usepackage{cite}
\usepackage{epsfig}
\renewcommand{\baselinestretch}{2}

\begin{document}

%
\title{Rich Magneto-absorption Spectra in AAB-stacked Trilayer Graphene \\}
\author{
\small Thi-Nga Do $^{a}$, Po-Hsin Shih $^{a}$, Cheng-Pong Chang $^{b}$, Chiun-Yan Lin$^{a,*}$, Ming-Fa Lin$^{a,*}$ $$\\
\small  $^a$Department of Physics, National Cheng Kung University, Tainan 701, Taiwan \\
\small $^b$Center for General Education, Tainan University of Technology, Tainan 701, Taiwan \\
 }
\renewcommand{\baselinestretch}{1}
\maketitle

\renewcommand{\baselinestretch}{1.4}
\begin{abstract}

The generalized tight-binding model is developed to investigate the feature-rich magneto-optical properties of AAB-stacked trilayer graphene. Three intragroup and six intergroup inter-Landau-level (inter-LL) optical excitations largely enrich the magneto-absorption peaks. In general, the former are much higher than the latter, depending on the phases and amplitudes of LL wavefunctions. The absorption spectra exhibit the single- or twin-peak structures which are determined by the quantum modes, LL energy spectra and Fermion distribution. The splitting LLs, with different localization centers (2/6 and 4/6 positions in a unit cell), can generate very distinct absorption spectra. There exist extra single peaks because of LL anti-crossings. AAB, AAA, ABA, and ABC stackings quite differ from one another in terms of the inter-LL category, frequency, intensity, and structure of absorption peaks. The main characteristics of LL wavefunctions  and energy spectra and the Fermi-Dirac function are responsible for the configuration-enriched magneto-optical spectra.

\vskip 1.0 truecm
\par\noindent

\noindent \textit{Keywords}: AAB-stacked trilayer graphene, magneto-absorption spectra, optical selection rules
\vskip1.0 truecm

\par\noindent  * Corresponding author.
{~ Tel:~ +886-6-275-7575.}\\~{{\it E-mail addresses}: l28981084@mail.ncku.edu.tw (C.Y. Lin), mflin@mail.ncku.edu.tw (M.F. Lin)}
\end{abstract}

\pagebreak
\renewcommand{\baselinestretch}{2}
\newpage

{\bf 1. Introduction}
\vskip 0.3 truecm

Few- and multi-layer graphenes with different stackings have attracted a lot of experimental \cite{ZYRong, JMCampanera, LYZhang, LBBiedermann, CTEllis, CHLui, JHHwang, ZYJuang, ZSWu} and theoretical \cite{YKHuang, YHLai, YPLin, SHRSena, CYLin, TNDo, YBZhang, CLKane, KSNovoselov, XJiang, JYWu} studies because of the special hexagonal symmetry.
They have been successfully synthesized by experimental methods, such as exfoliation of highly orientated pyrolytic graphite \cite{ZYRong, JMCampanera}, chemical and electrochemical reduction of graphene oxide \cite{LYZhang, LBBiedermann}, metalorganic chemical vapour deposition (MOCVD) \cite{CTEllis, CHLui, JHHwang, ZYJuang }, and arc discharge \cite{ZSWu}.
Graphene shows unusual essential physical properties, including the diverse optical selection rules \cite{YKHuang},$^{10}$ rich magnetic quantization \cite{YKHuang, YHLai, YPLin, SHRSena, CYLin, TNDo }, half-integer Hall effect \cite{LYZhang, YBZhang, CLKane , KSNovoselov}, low frequency plasma \cite{XJiang, JYWu}, and others.
Such properties could be easily modulated by stacking configurations \cite{CYLin, MAoki, AAAvetisyan, KFMak  }, dopping \cite{DMBasko, CCasiraghi}, layer number \cite{HHibino, EHMFerreira}, magnetic field ($\vec {\mathbf B } = B_0 \hat{z}$) \cite {YZhang, MOGoerbig}, electric field \cite{CLLu, EVCastro, KFMak2}, mechanical strain \cite{JELee, SMChoi, JHWong}, and temperature \cite{YWTan, VVCheianov  }.
To date, four important typical stacking configurations, AAB \cite{ZYRong, JMCampanera, LBBiedermann}, ABA \cite{LBBiedermann, CTEllis, CHLui }, AAA \cite{JHHwang, ZYJuang }, and ABC \cite{LBBiedermann, CHLui}, have been discovered.
In this work, we develop the generalized tight-binding model to investigate the magneto-absorption spectra of AAB-stacked trilayer graphene.
A detailed comparison among four stacking trilayer systems is also made.

The rich magnetic quantization is induced by the distinct stacking configurations.
A N-layer graphene possesses N groups of LLs, where each group consists of valence and conduction ones \cite{CYLin}.
Monolayer graphene has the $\sqrt B_0$-dependent inter-LL transition energies, in which the absorption peaks satisfy the selection rule of $\Delta n = \pm 1$ \cite{YHHo}.
Several theoretical studies on the magneto-optical properties have been elaborated for the AAA- and ABC-stacked trilayer graphenes.
The AAA stacking presents three pairs of Dirac cones results in only three intragroup inter-LL transition categories \cite{RBChen, CPChang}.
The above-mentioned $\sqrt B_0$ dependence and selection rules are also valid for N-layer AAA graphene.
The ABC stacking has three pairs of energy bands, including the linear, sombrero-shaped, and parabolic bands, which give rise to nine categories of inter-LL excitations \cite{YPLin2, SYuan}.
On the other hand, a thorough systematic study on the magneto-optical properties of the ABA-stacked trilayer graphene is absent up to now \cite{SYuan, SBerciaud }.
The electronic structure exhibits two pairs of parabolic bands and a pair of linear ones.
This means that the ABA stacking could be regarded as the combination of monolayer and AB-stacked bilayer graphenes in optical properties.
It is worth mentioning that the AB bilayer system presents four categories of absorption peaks, in which there is no simple relationship between the inter-LL transition energies and $B_0$ \cite{YHHo, YHHo2}.
The ABA graphene is thus expected to have five categories of inter-LL excitations, what will be investigated and presented in this work (Figs. 6(a)-6(c)).
The energy dispersions of AAB-stacked trilayer graphene consist of three divergent pairs of valence and conduction bands, oscillatory ($S_1^{c,v}$), sombrero-shaped ($S_2^{c,v}$), and parabolic ones ($S_3^{c,v}$) (Fig. 1(b)) \cite{TNDo}.
Significantly, the $S_1^{c,v}$ bands (Fig. 1(c)) are first revealed in the AAB stacking implying that its electronic structure is exclusive among the aforementioned trilayer systems.
How the unique electronic properties reflected in the magneto-absorption spectra is worthy of a thorough study.

The methods of optical spectroscopies are powerful tools to identify the optical excitations and selection rules in condensed-matter systems.
Such experimental techniques, including absorption, transmission, reflection, Raman scattering and Rayleigh scattering spectroscopies, have been successfully applied to two-dimensional (2D) materials, concerning a large class of graphene-related systems, such as, few-layer graphenes \cite{CCasiraghi2}, carbon nanotubes \cite{SMTabakman}, graphene nanoribbons \cite{RDenk}, graphite and graphite intercalation compound \cite{YKim}; also other few-layer systems, e.g., silicenes \cite{DJose, EScalise  }, topological insulators \cite{JZhang}, $MoS_2$ \cite{BCWindom , HLi}, and so on.
The intra- and inter-LL optical transitions in graphene systems have been observed in infrared spectroscopy.
Evidently, the field-dependent frequency and the selection rule of $\Delta n = \pm 1$ are identified by transmission and Raman scattering spectroscopy \cite{EHenriksen, ZJiang, PPlochocka, CFaugeras, HZhao, CCong}.
In particular, the square-root and linear $B_0$-dependent frequencies due to the massless and massive Dirac-fermions, respectively, of the monolayer and AB-stacked bilayer graphenes, are revealed in the infrared transmission spectra \cite{EHenriksen, ZJiang, PPlochocka}.
These Dirac-fermions are further confirmed by the magneto-Raman spectroscopy for the few-layer AB-stacked and ABC-stacked graphenes \cite{CFaugeras, HZhao, CCong}.
It should be noted that the experimental measurements mentioned previously are only resolved on the intragroup magneto-absorption spectra due to the first group of LLs.

The center of interest in this work is to investigate the configuration-enriched magneto-optical properties of AAB-stacked trilayer graphene using the generalized tight-binding model.
The special electronic properties are directly reflected on the LL energies and wavefunctions, so that they induce the extraordinary features in the magneto-optical properties.
This study shows that there exist nine categories of absorption peaks, including three intragroup and six intergroup ones.
The intragroup transitions remarkably differ from the intergroup ones in the intensity, energy spacing, and structure of peaks.
The former exhibits very low intensity, while the latter are much higher and comparable with those of monolayer graphene.
Moreover, the splitting LLs, arising from the destruction of inversion symmetry, can diversify the magneto-absorption spectra.
These characteristics are absent in the other trilayer graphene systems.
There are important differences among four typical stacking trilayer graphenes in the excitation category and the main features of absorption peaks.
The predicted results, with regard to nine categories of inter-LL transitions, can be verified by optical spectroscopies \cite{EHenriksen, ZJiang, PPlochocka, CFaugeras, HZhao, CCong}.
\newpage
\vskip 0.6 truecm
\par\noindent
{\bf 2. Method }
\vskip 0.3 truecm

The generalized tight-binding model is developed to study the magneto-electronic spectra of AAB-stacked trilayer graphene \cite{TNDo}.
For the first two layers, shown in Fig. 1(a), all carbon atoms have the same (x,y) projections while the third layer can be obtained by shifting the first (or the second) one by a distance of $b$ along the armchair direction.
The parameters of ten intra- and inter-layer atomic interactions are used ($\gamma_{0}=-2.569  $ eV, $ \gamma_{1}=-0.263$ eV, $\gamma_{2}=0.32 $ eV, $ \gamma_{3}=-0.413$ eV, $\gamma_{4}=-0.177 $ eV, $ \gamma_{5}=-0.319$  eV, $ \gamma_{6}=-0.013  $ eV, $ \gamma_{7}=-0.0177 $ eV, $ \gamma_{8}=-0.0319  $ eV, and $ \gamma_{9}=-0.012 $ eV ) to characterize the special energy bands.
A uniform perpendicular magnetic field can induce the Peierls phase related to the vector potential and thus enlarge the unit cell to become a rectangle including 12$R_B$ carbon atoms ($R_B=hc/eB_0$) (Fig. 1(a)).
The eigenfunctions of the quite large Hamiltonian is solved by an exact diagonalization method to be
 in the arrangement of $12R_B$ tight-binding functions as follows.
\begin {equation*}
|\Psi_{\mathbf{k}}^{c,v}\rangle = \sum_{l=1}^{3}\sum_{m=1}^{2R_{B}} \bigg ( c_{A_{m,\mathbf{k}}^l }\bigg| A_{m,\mathbf{k}}^l \bigg \rangle    + c_{B_{m,\mathbf{k}}^l } \bigg| B_{m,\mathbf{k}}^l \bigg \rangle \bigg ) ,
\end {equation*}
in which $c$ and $v$ denote, respectively, the conduction and valence states; $ c_{A_{m,\mathbf{k}}^l }$ ($ c_{B_{m,\mathbf{k}}^l }$) represent the subenvelope functions on the six distinct sublattices.
For each LL, the quantum number ($n_l^{c,v}$) could be determined by the number of zero points in the dominating subenvelope wavefunctions.

When ABA-stacked trilayer graphene is present in an electromagnetic field with the electric polarization $\hat{E} \parallel \hat{y}$, an occupied LL will be excited to an unoccupied one.
The optical absorption function, according to the Fermi golden rule, is expressed as
\begin {align}
A(\omega) \propto
&\sum_{c,v,m,m'} \int_{1stBZ} \frac {d\mathbf{k}}{(2\pi)^2}
 \Big| \Big\langle \Psi^{c} (\mathbf{k},m')
 \Big| \frac{   \hat{\mathbf{E}}\cdot \mathbf{P}   } {m_e}
 \Big| \Psi^{v}(\mathbf{k},m)    \Big\rangle \Big|^2 \nonumber \\
 &\times
Im \Big[      \frac{f(E^c (\mathbf{k},m')) - f(E^v (\mathbf{k},m))}
{E^c (\mathbf{k},m')-E^v (\mathbf{k},m)-\omega - i\Gamma}           \Big] ,
\end {align}
 where $ \mathbf{P}$ is the momentum operator, $f(E^{c,v} (\mathbf{k},m))$ the Fermi-Dirac distribution function; $\Gamma$ ($\approx 1$ meV) the broadening parameter.
The absorption spectrum is associated with the velocity matrix elements (the first term) and the joint density of states (the second term).
The former can determine whether the inter-LL transitions are available.
The velocity matrix elements, as successfully done for carbon-related materials \cite{MFLin}, are evaluated under the gradient approximation in the form of
\begin{align*}
\Big\langle \Psi^{c} (\mathbf{k},m')
 \Big| \frac{   \hat{\mathbf{E}}\cdot \mathbf{P}   } {m_e}
 \Big| \Psi^{v}(\mathbf{k},m) \Big\rangle
&\cong \dfrac{ \mathrm{\partial}}{\mathrm{\partial}k_y  }
\Big\langle \Psi^{c} (\mathbf{k},m')  \Big| H \Big|  \Psi^{v}(\mathbf{k},m) \Big\rangle \\
&= \sum_{l,l' = 1}^{3} \sum_{m,m' = 1}^{2R_B}
\bigg (
c_{A_{m,\mathbf{k}}^{l}}^*  c_{A_{m',\mathbf{k}}^{l'}}
\dfrac{ \mathrm{\partial}}{\mathrm{\partial}k_y  }
 \Big\langle A_{m,\mathbf{k}}^{l} \Big| H \Big|  A_{m,\mathbf{k'}}^{l'}  \Big\rangle \\
&+ c_{A_{m,\mathbf{k}}^{l}}^*  c_{B_{m',\mathbf{k}}^{l'}}
\dfrac{ \mathrm{\partial}}{\mathrm{\partial}k_y  }
 \Big\langle A_{m,\mathbf{k}}^{l} \Big| H \Big|  B_{m,\mathbf{k'}}^{l'}  \Big\rangle \\
&+ c_{B_{m,\mathbf{k}}^{l}}^*  c_{A_{m',\mathbf{k}}^{l'}}
\dfrac{ \mathrm{\partial}}{\mathrm{\partial}k_y  }
 \Big\langle B_{m,\mathbf{k}}^{l} \Big| H \Big|  A_{m,\mathbf{k'}}^{l'}  \Big\rangle \\
&+ c_{B_{m,\mathbf{k}}^{l}}^*  c_{B_{m',\mathbf{k}}^{l'}}
\dfrac{ \mathrm{\partial}}{\mathrm{\partial}k_y  }
 \Big\langle B_{m,\mathbf{k}}^{l} \Big| H \Big|  B_{m,\mathbf{k'}}^{l'}  \Big\rangle
\bigg ).
\end {align*}
Since the nearest-neighbor atomic interaction is much larger than the interlayer ones, only the Hamiltonian elements of the former dominate the velocity matrix elements.
On the same layer, $\mathrm{\partial}H/\mathrm{\partial}k_y$ is almost the same for all the nearest neighbors in the enlarged unit cell.
Consequently, the magneto-optical selection rule depends on the relations between the initial state on the $A^l$ ($B^l$) sublattice and the final state on the $B^l$ ($A^l$) one.
That is, these two states on the different sublattices of the same layer must have the same quantized mode.

\vskip 0.6 truecm
\par\noindent
{\bf 3. Magneto-absorption spectra}
\vskip 0.3 truecm

AAB-stacked trilayer graphene exhibits three groups of LLs, where each group consists of occupied valence and unoccupied conduction ones.
Each LL is characterized by the subenvelope functions on the six sublattices, as indicated in Figs. 2 and 3 for the $(k_x=0, k_y=0)$ state at 2/6 localization center under $B_0 =$ 40 T.
The first, second and third groups are initiated at 0 (0), -0.23 (0.24) and -0.58 (0.55) eV for the valence (conduction) states, as indicated by blue, green and red colors, respectively.
The valence and conduction LLs are asymmetric about $E_F=0$ because of the interlayer atomic interactions.
It is expected to induce the twin-peak structures in the magneto-absorption spectra.
The quantum numbers of the first ( $n_1^{c,v}$), second $(n_2^{c,v})$ and third groups $(n_3^{c,v})$ are, respectively, determined by the $A^1$, $B^2$ and $A^2$ sublattices with the dominating subenvelop functions (Figs. 2 and 3).
As to the first group, there exist three LLs near $E_F$, in which the middle $n_1^{c,v}=0$ LL is located right at $E_F$, while the other two are assigned $n_1^{c,v}=1$.
The next LLs $n_1^{c,v}=2$, $n_1^{c,v}=3$, ... are in an usual monolayer-like ordering.
Distinctly, the quantum numbers of $n_2^{c,v}$ begin from $n_2^{c,v} = 1$, then $n_2^{c,v} = 0$, $n_2^{c,v} = 2$, $n_2^{c,v} = 3$, and so on.
As for $n_3^{c,v}$, it normally increases from zero for the higher conduction states, and the lower valence states.

The stacking configuration of AAB results in the special relationship between the different subenvelop functions.
All the LL wavefunctions are almost well-behaved in the spatial distributions at $B_0=$ 40 T, in which the main modes are much stronger than the side ones.
The subenvelope functions of $A^1$ and $A^2$ ($B^1$ and $B^2$) are similar in the number of zero points since the first and second layers have the same (x,y) projection \cite{TNDo}.
That is to say, the quantum numbers determined by the $A^1$ and $A^2$ sublattices are identical.
It should be noted that, for the first and second groups (the third one), the phases of the $A^1$ and $A^2$ sublattices are identical for all the conduction (valence) states but opposite for the valence (conduction) ones.
Moreover, the spatial distribution of certain subenvelope functions are much smaller compared to the others.
For example, for the first group, the amplitude of the $A^3$ sublattic is much weaker than those of the other sublattices (Fig. 2).
The critical differences in amplitude and phase are expected to induce the unique absorption spectra.

The nine categories of optical excitations, which include three intragroup and six intergroup ones, are clearly revealed in the magneto-absorption spectrum (Figs. 4(a)-4(e)).
The inter-LL optical excitations between any two groups are available only under the requirement of a specific selection rule.
The optical absorption peaks corresponding to the first, second, and third LL groups are respectively indicated by black, green, and red numbers (underline for valence LLs).
The intragroup and intergroup absorption peaks quite differ from each other in the spectral intensities, peak structures, and energy spacings in twin peaks.
The whole intragroup categories, excluding the threshold peak $\color{blue}{\underline{0}1}$, are twin peaks with very low intensity.
While for the intergroup ones, except for the transitions between the first and third LL groups, all of peaks show the single-peak structure with monolayer-like intensity.
The intensity difference arises from the special relations of the subenvelope functions between the $A^l$ ($B^l$) sublattice of the initial states and the $B^l$ ($A^l$) sublattice of the final ones (discussed later).
Regarding the energy spacings in twin peaks, those of the intergroup excitations are much wider than the intragroup ones, mainly owing to the asymmetry of LL energy spectra.

The intensities of absorption peaks are determined by the summation of $\langle A^l | B^l \rangle$ and $\langle B^l | A^l \rangle$, where $l = 1, 2, 3$.
As to the first and third groups, $\langle A^3 | B^3 \rangle \approx 0$, especially for low-lying LLs because of the very low amplitudes of the $A^3$ and $B^3$ subenvelope functions, respectively.
Besides, the contributions of $\langle A^2 | B^2 \rangle$ are fainter for higher or deeper LLs due to the lower amplitudes on two sublattices.
Moreover, the peak intensity can be largely enhanced or reduced depdending on the phase relationship between the two sublattices: $A^l$ ($B^l$) of the initial states and $B^l$ ($A^l$) of the final ones.
In general, the intensities of the intergroup transition peaks are much higher than those of the intragroup ones.
For example, the intragroup peak $\color{blue}\underline{1}2$ at $\omega\approx 0.06$ eV (marked by blue arrow) is about 1/15 compared with the intergroup one $\color{green}\underline{1} \color{blue}1$ at $\omega\approx 0.25$ eV (marked by green arrow in Fig. 4(a)).
For the former, the $|B^1\rangle$ of the $n_1^v=1$ initial state and the $\langle A^1|$ of the $n_1^c=2$ final state have the distinct phases, while $|B^2\rangle$ and $\langle A^2|$ possess the same ones.
As a result, $\langle A^1 | B^1 \rangle$ and $\langle A^2 | B^2 \rangle$ present the opposite signs which reduce the velocity matrix elements, and thus lower the absorption peak intensity.
On the contrary, the peak intensity of the latter is enhanced due to the fact that all the dominating subenvelope functions of the $n_2^v=1$ initial and $n_1^c=1$ final states on the sublattices, ($B^1$, $A^1$), ($B^2$, $A^2$); ($A^3$, $B^3$), have similar phase relations.
It should be noted that there are a number of extremely low peaks in the spectra, e.g., $\color{blue}\underline{5}1$, $\color{blue}\underline{5}1$ (red arrow), and so on, due to the appearance of the side modes.
Those transitions do not satisfy the optical selection rule and thus could be neglected.

The intragroup inter-LL excitations possess a lot of low-intensity absorption peaks, as shown in Fig. 4(a).
The threshold peak of $n_1^v n_1^c= \color{blue}\underline{0}1$, with frequency of $\omega=0.01$ eV, possesses an ignorable intensity (inserted in Fig. 4(a)).
This is caused by the extremely weak subenvelope functions of $n_1^v=0$ LL in the $B^1$, $A^2$ and $A^3$ sublattices.
The other low peaks belong to the twin-peak structures, such as $n_1^v n_1^c= \color{blue}\underline{2}1$ and $n_1^v n_1^c= \color{blue}\underline{1}2$ ($\omega\approx 0.05$ eV), $n_1^v n_1^c= \color{blue}\underline{3}2$ and $n_1^v n_1^c= \color{blue}\underline{2}3$ ($\omega\approx 0.13$ eV), and so on.
The quantum mode of the $n_1^v=2$ $A^1 (A^2)$ sublattice is identical to that of the $n_1^c=1$ $B^1 (B^2)$ one, so that the $\color{blue}{\underline{2}1}$ peak satisfies the optical selection rule.
However, the same phase between $A^1$ and $B^1$ sublattices competes with the opposite phases of the $A^2$ and $B^2$ ones, which predominantly reduces the peak intensity.
Concerning the ($n_2^v \rightarrow n_2^c$) and ($n_3^v \rightarrow n_3^c$) excitation channels, the ($n_2^v n_2^c= \color{green}\underline{0}1$, $n_2^v n_2^c= \color{green}\underline{1}0$) and ($n_3^v n_3^c= \color{red}\underline{0}1$, $n_3^v n_3^c= \color{red}\underline{1}0$) threshold peaks are, respectively, revealed at ($\omega=0.502$ eV; $\omega=0.51$ eV ) and ($\omega= 1.205$ eV; $\omega= 1.209$ eV).
While the intensities of the latter two are comparable with the other intragroup excitation peaks, those of the two formers are predominant.
This is because all the subenvelope functions of the $n_2^v=0$ (1) initial and $n_1^c=1$ (0) final states on the $A^m$ and $B^m$ sublattices have the same (opposite) phases.
In general, three categories of intragroup inter-LL excitations exhibit similar characteristics, including peak intensity, twin-peak structure, and selection rule.

The intergroup inter-LL excitations exhibit single and twin peaks, depending on the quantum modes and the asymmetry of LL energy spectra.
The absorption categories related to the second group of LLs, the $n_1^v \rightarrow n_2^c$, $n_2^v \rightarrow n_1^c$, $n_3^v \rightarrow n_2^c$, and $n_2^v \rightarrow n_3^c$ transitions, present solely single-peak structures.
For the two formers, the initial and final states of available excitations satisfy $n_1^{c,v} = n_2^{v,c}$, e.g., $ \color{blue}{\underline 0} \color{green}0$, $ \color{green}{\underline 0} \color{blue}0$, $ \color{blue}{\underline1} \color{green}1$, $ \color{green}{\underline1} \color{blue}1$, and others (Fig. 4(a)).
Especially, their absorption spectra begin, respectively, with the peaks of $ n_1^v n_2^c = \color{blue}{\underline1} \color{green}1$ ($\omega = 0.25$ eV; marked by green arrow) and $n_2^v n_1^c=\color{green}{\underline1} \color{blue}{1}$ ($\omega = 0.245$ eV), according to the unusual ordering of quantum numbers in the second group (Fig. 3).
And then the threshold peaks, $ n_1^v n_2^c = \color{blue}{\underline0} \color{green}0$ and $n_2^v n_1^c=\color{green}{\underline0} \color{blue}{0}$, are merged together at $\omega = 0.265$ eV and become a higher-intensity single peak.
Concerning the $n_2^v \rightarrow n_3^c$ and $n_3^v \rightarrow n_2^c$ excitations, the inter-LL transitions with $| n_2^{c,v} - n_3^{v,c}| = 2$ and 0 can induce the single-peak structure.
The former are much higher than the latter, mainly owing to the special relations among the subenvelope functions.
For instance, the $\color{green}{\underline2} \color{red}0$ single peak satisfies the optical selection rule where the quantum modes in the $A^1, A^2, B^3 $ sublattices of $n_2^v=2$ are, respectively, identical to those in the $B^1, B^2$, and $A^3$ sublattices of $n_3^c=0$.
Furthermore, the corresponding sublattices with the same phases are responsible for the predominant peak intensity, compared to the weak peaks of $\color{green}{\underline0} \color{red}0$, $\color{green}{\underline1} \color{red}1$, and so on.
On the other hand, the $n_3^{v}\rightarrow n_1^{c}$ and $n_1^{v}\rightarrow n_3^{c}$ excitations exhibit the twin-peak structures.
This is due to the fact that the inter-LL transitions obey the selection rule of $| n_1^{c,v} - n_3^{v,c}| = 1$, similar to that of the intragroup ones.
The twin peaks possess very different intensities in which one resembles the intragroup peaks while another is comparable to the intergroup ones.
For example, the $\color{blue}{\underline2} \color{red}1$ and $\color{blue}{\underline1} \color{red}{2}$ peaks are, respectively, weak and strong as indicated by red arrows in Fig. 4(b).
Exceptionally, the threshold peak $\color{red}{\underline0} \color{blue}1$ is a single one since the $\color{red}{\underline1} \color{blue}0$ transition is forbidden by the Fermi-Dirac distribution function.

The splitting of LLs results in critical differences between magneto-optical absorption spectra corresponding to 4/6 (Figs. 5(a)-5(e)) and 2/6 localization centers in terms of intensity, frequency, and structure of peaks. This is mainly owing to the abnormal quantum number ordering and the significant side modes during the LL anti-crossings \cite{TNDo}. Accordingly, the dissimilarities of two spectra are clearly revealed in the inter-LL excitations related to the low-lying LLs. For the threshold peak $\color{blue}{\underline0 1}$, that of the former has much higher intensity and frequency ($\omega \approx 0.15$ eV) (Fig. 5(a)). There are a lot of low-frequency single peaks when the magnetic field is within the range of LL anti-crossings. For example, the $n_1^{c,v}=0$ LL also has the non-negligible side mode of three zero points, being responsible for the anti-crossing of the $n_1^{c,v}=0$ and $n_1^{c,v}=3$ LLs and the appearance of the single peaks $\color{blue}{\underline0 2}$ and $\color{blue}{\underline0 4}$. Furthermore, many pairs of LLs in the anti-crossings such as ($n_1^{c,v}=1$, $n_1^{c,v}=4$), ($n_1^{c,v}=2$, $n_1^{c,v}=5$), and so on, will present several extra single peaks. Consequently, similar characteristics also occur in the initial absorption peaks of the intergroup transitions between the first LL group and others. Different from the spectrum at 2/6 center, the $n_2^v \rightarrow n_3^c$ and $n_3^v \rightarrow n_2^c$ excitations only present the inter-LL transitions with $| n_2^{c,v} - n_3^{v,c}| = 0$. In short, the LL anti-crossings play an important role in distinguishing the magneto-absorption spectra due to the splitting LLs.

 Specially, the magneto-absorption spectrum of ABA-stacked trilayer graphene is worthy of a detailed consideration for the wide-frequency range. Since LLs are four-fold degenerate, the inter-LL excitations at 4 different localization centers, 1/6, 2/6, 4/6, and 5/6, are identical. Only five categories of absorption peaks are revealed in the absorption spectrum, as shown in Figs. 6(a)-6(c). They correspond to those of AB-stacked bilayer \cite{YHHo,YHHo2} and monolayer graphenes \cite{YHHo}, mainly owing to the absence of the inter-LL transitions related to a pair of linear bands and two pairs of parabolic ones. Two categories of intergroup excitations, $2^v \rightarrow 3^c$ and $3^v \rightarrow 2^c$, exhibit twin peaks due to the asymmetric LL energy spectrum. The intragroup inter-LL excitations of the second and third LL groups are mostly present in twin-peak structures except for the threshold $\color{green}{\underline{0} 1}$ peak. The absence of $\color{green}{\underline{1} 0}$ transition is due to the fact that the $n=0$ LL is occupied. Especially, the category of $1^v \rightarrow 1^c$ transitions generally displays single peaks with dominant uniform intensity because of the symmetric LL energy spectrum of the quantized linear bands. There is only a twin-peak structure for $\color{blue}{\underline{2} 1}$ and $\color{blue}{\underline{1} 2}$ (red arrows in Fig. 6(b)), reflecting the distortion of the linear bands near the Dirac points. Moreover, the peak frequencies are proportional to $\sqrt{B_0}$, similar to those in three absorption categories of the AAA stacking. On the other hand, there is no simple relationship in the other inter-LL excitations of the ABA, ABC and AAB stackings.

The stacking configurations can diversify the magnetic quantization and thus the magneto-absorption spectra. Four typical trilayer graphenes, AAB, AAA, ABC, and ABA stackings, are quite different from one another in the excitation category, structure, intensity, and number of absorption peaks. The differences mainly come from the special relations among the subenvelope functions on the six sublattices. The AAA trilayer graphene, being regarded as three separated monolayer graphenes, only has three categories of intragroup absorption peaks \cite{RBChen, CPChang}.
The main reason for the absence of the intergroup excitations is that all the LL wavefunctions are the linear symmetric or asymmetric superpositions of the subenvelope functions in different layers. The absorption spectrum only exhibits the single-peak structures with uniform intensity except for the threshold peak in each category with half of intensity. On the other hand, the ABC-stacked trilayer graphene possesses nine categories of absorption peaks with nonuniform intensities \cite{YPLin2, SYuan}, similar to those of the AAB stacking. The whole spectrum, excluding the threshold peaks, exhibits the twin-peak structure. Overall, the intragroup excitations are stronger than the intergroup ones. This can also be observed in the ABA stacking (Fig. 6(a)-6(c)), being in sharp contrast with the AAB stacking. Concerning the number of absorption peaks, it is higher for the ABC and AAB stackings but lowest for the AAA one. The critical differences among the four stacked trilayer graphenes can provide the useful information in identifying the specific stacking configurations by the experimental measurements \cite{EHenriksen, ZJiang, PPlochocka, CFaugeras, HZhao, CCong}.

\vskip 0.6 truecm
\par\noindent
{\bf 4. Conclusion }
\vskip 0.3 truecm

The magneto-optical properties of the AAB-stacked trilayer graphene have been investigated using the generalized tight-binding model.
This method can also be extended to study the other essential physical properties of few-layer graphenes and other 2D materials.
An optical selection rule that the $A^l$ ($B^l$) sublattice of the initial state has the same quantum mode with the $B^l$ ($A^l$) sublattice of the final state is deduced from the spatial distributions of LL wavefunctions.
The main features of magneto-absorption spectra, including the excitation category, intensity, structure, frequency, and number of absorption peaks, are fully explored in a wide frequency range.
There are nine categories of inter-LL optical transitions in which the intragroup and intergroup ones quite differ from one another.
The magneto-absorption spectrum could be enriched by the splitting LLs, arising from the destruction of inversion symmetry in the AAB stacking.
Both single- and twin-peak structures are clearly revealed in the magneto-absorption spectra.
The main characteristics of LL wavefunctions  and energy spectra and the Fermi-Dirac function are responsible for the above-mentioned characteristics. The predicted feature-rich magneto-absorption spectra with nine excitation categories can be further verified by optical spectroscopy techniques \cite{EHenriksen, ZJiang, PPlochocka, CFaugeras, HZhao, CCong}.

The intergroup and intragroup absorption peaks are significantly different in intensity, structure, and frequency.
 In general, the former are much higher than the latter, depending on the phase and amplitude relations between the subenvelope functions on six sublattices.
Due to the quantum modes and asymmetric LLs, the former have single and twin peaks while the latter display twin-peak structures except for threshold ones.
The energy spacings between the twin-peak structures are much wider for the intergroup excitations compared with the intragroup ones.
The critical differences in spectra at 2/6 (5/6) and 4/6 (1/6) localization centers mostly lie in the excitations associated to the low-lying LLs.
This is caused by the significant side modes during the LL anti-crossings.
A lot of low-frequency extra single peaks are present in the latter when the magnetic field is within the range of LL anti-crossings.
The inter-LL excitations of the higher-lying LLs are quite similar in certain respects for different localization centers.

The AAB-stacked trilayer graphene exhibits the unique magneto-optical properties among four typical stacking systems. It possesses the most plentiful magneto-absorption spectrum, mainly owing to the splitting LLs and the frequent LL anti-crossings. The AAA stacking, the superposition of three monolayer graphenes, has three categories of intragroup excitation peaks. The magneto-absorption spectrum only has many single peaks with a uniform intensity except for the threshold ones. The ABA stacking, being regarded as the combination of monolayer and AB-stacked bilayer graphene, only presents five categories of absorption peaks. Only peaks arising from the slightly distorted Dirac cones show the uniform spectral intensity in the single-peak structures. Two pairs of parabolic bands induce two intragroup and two intergroup excitations categories, in which the former are stronger than the latter. All the peaks, the threshold ones excepted, have the twin-peak structures. The ABC stacking possesses the same number of intra- and inter-group absorption peak categories as the AAB one. However, the intensity relation between the intragroup and intergroup peaks is similar to that of the ABA stacking, but in sharp contrast with the AAB one.

\par\noindent {\bf Acknowledgments}

This work was supported in part by the National Science Council of Taiwan,
the Republic of China, under Grant Nos. NSC 98-2112-M-006-013-MY4 and NSC 99-2112-M-165-001-MY3.

\newpage
\renewcommand{\baselinestretch}{0.2}

\newpage \centerline {\Large \textbf {FIGURE CAPTIONS}}

Fig. 1 - The interlayer atomic interactions and the geometric structure under a uniform magnetic field $B_0\hat{z}$ (a). The shaded region corresponds to a rectangular unit cell. The first and second layers have the same (x, y) projections. (b) The low-energy band structures of AAB-stacked trilayer graphene with (c) a narrow energy gap. The solid and dashed curves represent the results by the use of the tight-binding model and the first-principle calculation, respectively.

Fig. 2 - The distributions of the low-lying LL wavefunctions of AAB-stacked trilayer graphene with the six distinct sublattices centered at the 2/6 localization under $B_0$ = 40 T. The unit of the x-axis is $m/2R_B$, where $m$ represents the $m$-th A or B atom in the enlarged unit cell.

Fig. 3 - The distributions of the conduction (a) and valence (b) LL wavefunctions of AAB-stacked trilayer graphene with the six distinct sublattices centered at the 2/6 localization under $B_0$ = 40 T. The unit of the x-axis is $m/2R_B$, where $m$ represents the $m$-th A or B atom in the enlarged unit cell.

Fig. 4 - Magneto-absorption spectra of AAB-stacked trilayer graphene in (a), (b), (c), (d) and (e) near 2/6 center with respect to different frequency ranges at $B_0=40$ T. The numbers with and without the underline stand for the quantum numbers of the occupied and unoccupied LLs, respectively. The blue, green, and red colors correspond to the first, second, and third groups of LLs.

Fig. 5 - Magneto-absorption spectra of AAB-stacked trilayer graphene in (a), (b), (c), (d) and (e) near 4/6 center with respect to different frequency ranges at $B_0=40$ T. The numbers with and without the underline stand for the quantum numbers of the occupied and unoccupied LLs, respectively. The blue, green, and red colors correspond to the first, second, and third groups of LLs.

Fig. 6 - Magneto-absorption spectra of AAB-stacked trilayer graphene in (a), (b), and (c) near 2/6 center with respect to different frequency ranges at $B_0=40$ T. The numbers with and without the underline stand for the quantum numbers of the occupied and unoccupied LLs, respectively. The blue, green, and red colors correspond to the first, second, and third groups of LLs.




\vskip0.5 truecm


\begin{figure}[htb]
\centering
\includegraphics[width=0.9\linewidth]{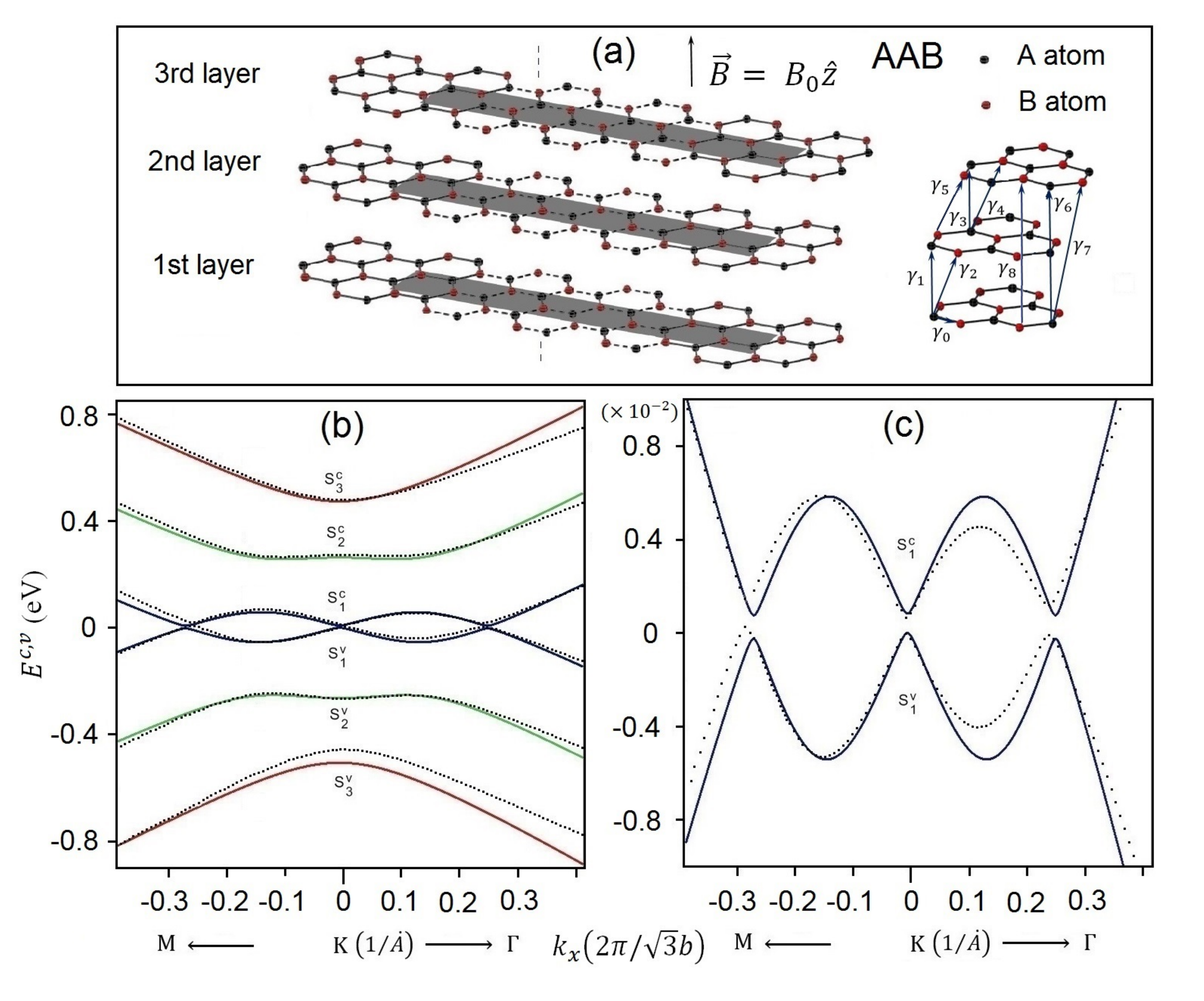}
\caption{The interlayer atomic interactions and the geometric structure under a uniform magnetic field $B_0\hat{z}$ (a). The shaded region corresponds to a rectangular unit cell. The first and second layers have the same (x, y) projections. (b) The low-energy band structures of AAB-stacked trilayer graphene with (c) a narrow energy gap. The solid and dashed curves represent the results by the use of the tight-binding model and the first-principle calculation, respectively.}
\end{figure}

\begin{figure}[htb]
\centering
\includegraphics[width=0.9\linewidth]{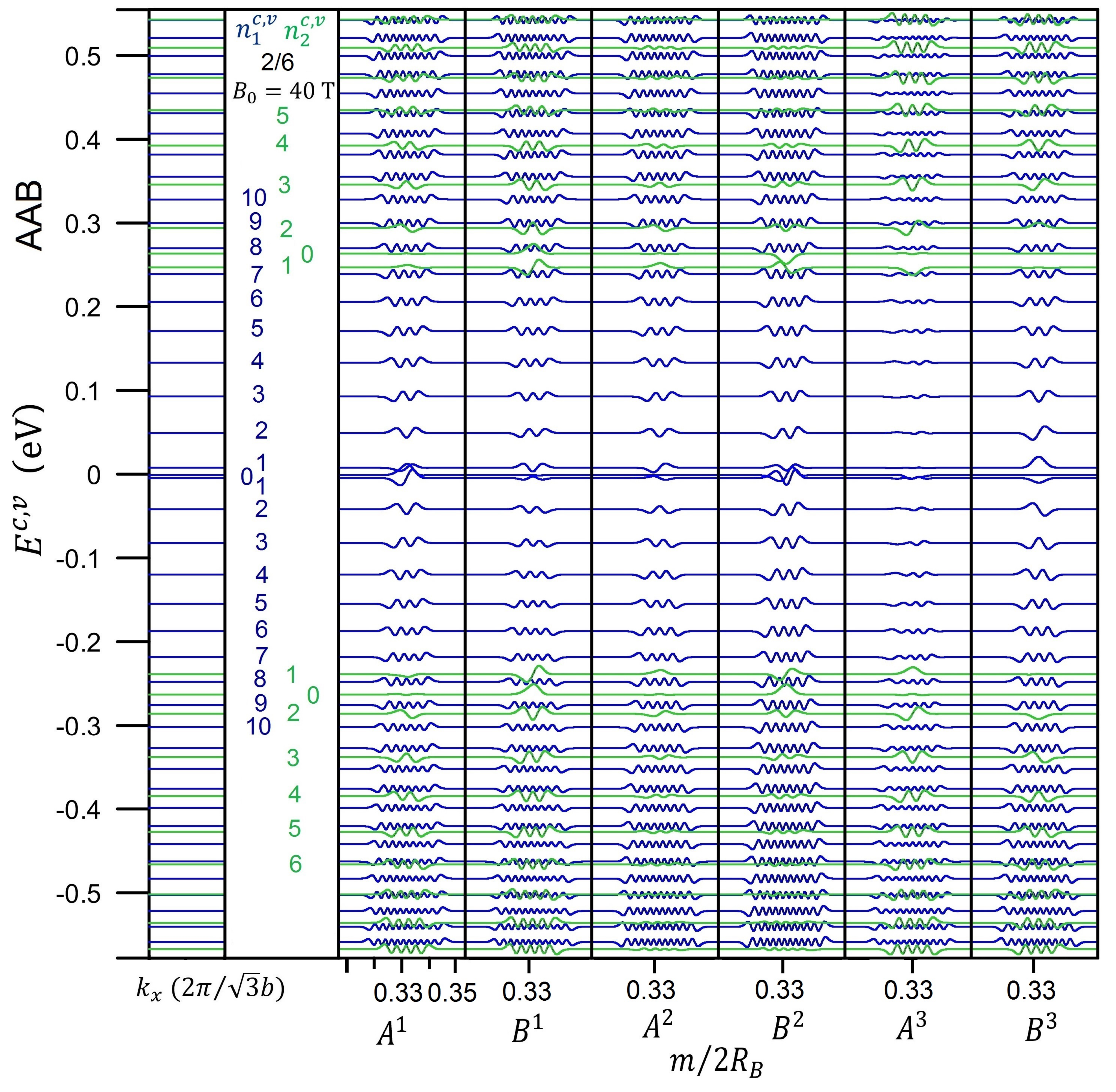}
\caption{The distributions of the low-lying LL wavefunctions of AAB-stacked trilayer graphene with the six distinct sublattices centered at the 2/6 localization under $B_0$ = 40 T. The unit of the x-axis is $m/2R_B$, where $m$ represents the $m$-th A or B atom in the enlarged unit cell.}
\end{figure}

\begin{figure}[htb]
\centering
\includegraphics[width=0.9\linewidth]{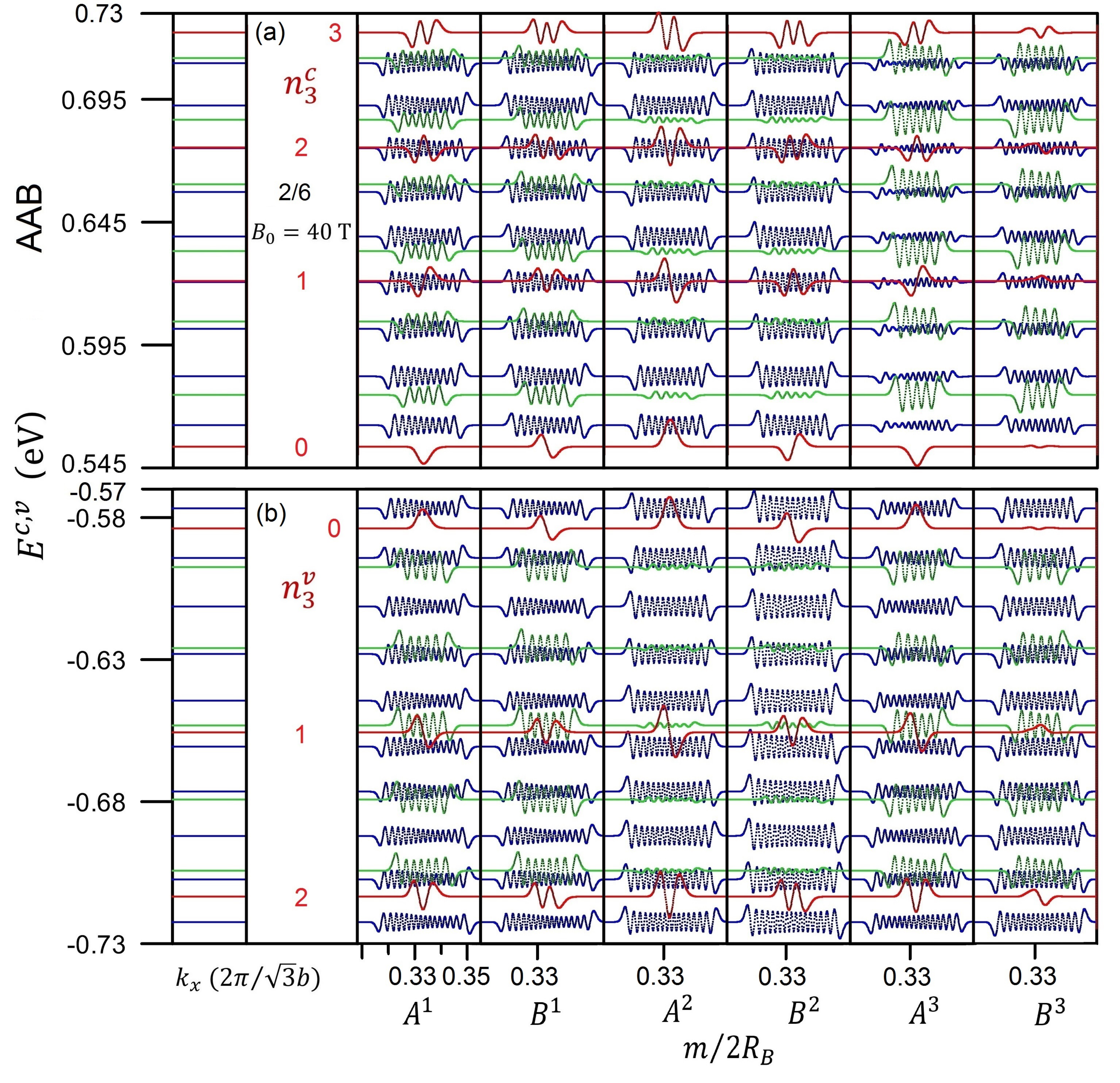}
\caption{The distributions of the conduction (a) and valence (b) LL wavefunctions of AAB-stacked trilayer graphene with the six distinct sublattices centered at the 2/6 localization under $B_0$ = 40 T. The unit of the x-axis is $m/2R_B$, where $m$ represents the $m$-th A or B atom in the enlarged unit cell.}
\end{figure}

\begin{figure}[htb]
\centering
\includegraphics[width=0.9\linewidth]{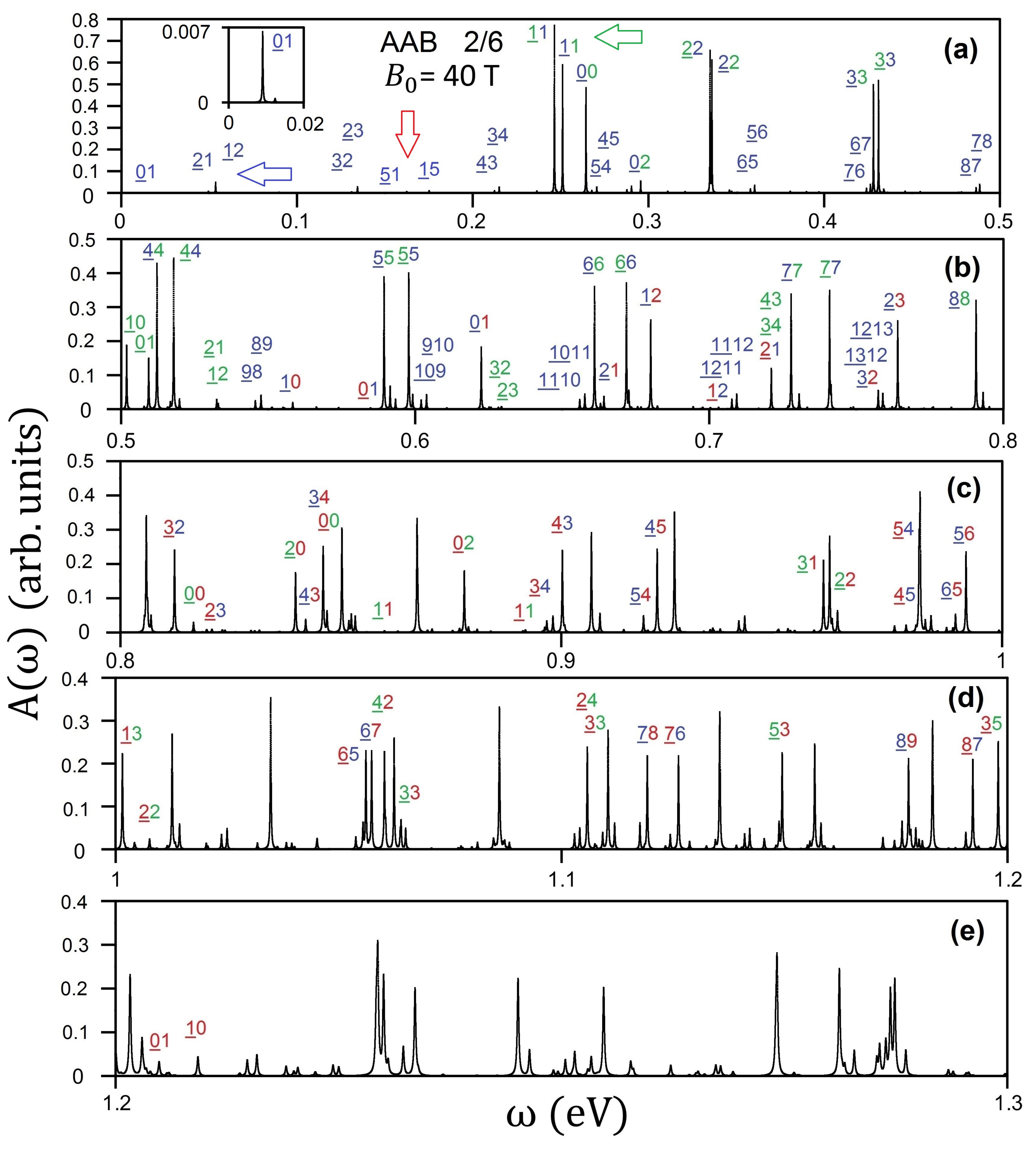}
\caption{Magneto-absorption spectra of AAB-stacked trilayer graphene in (a), (b), (c), (d) and (e) near 2/6 center with respect to different frequency ranges at $B_0=40$ T. The numbers with and without the underline stand for the quantum numbers of the occupied and unoccupied LLs, respectively. The blue, green, and red colors correspond to the first, second, and third groups of LLs.}
\end{figure}

\begin{figure}[htb]
\centering
\includegraphics[width=0.9\linewidth]{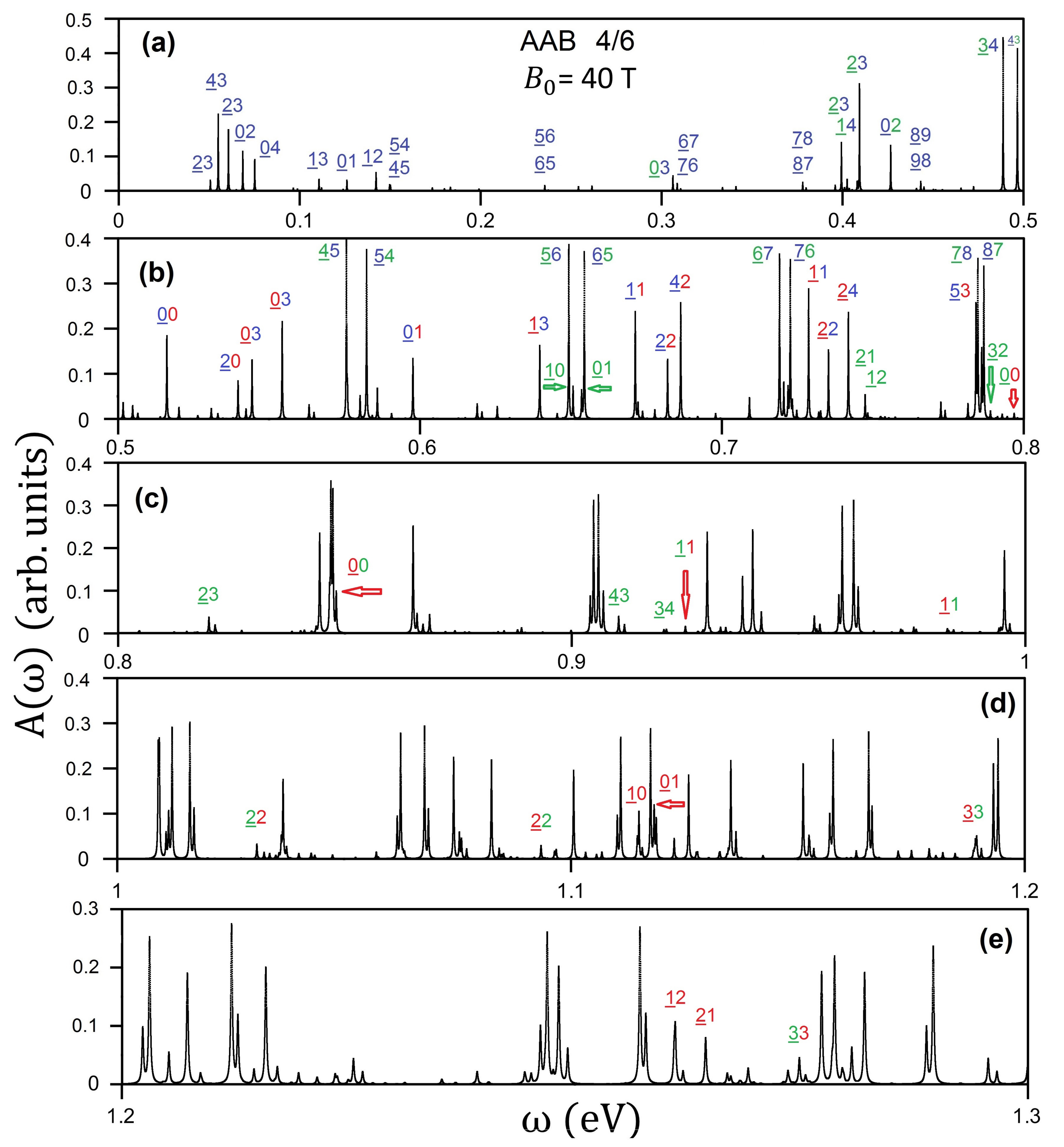}
\caption{Magneto-absorption spectra of AAB-stacked trilayer graphene in (a), (b), (c), (d) and (e) near 4/6 center with respect to different frequency ranges at $B_0=40$ T. The numbers with and without the underline stand for the quantum numbers of the occupied and unoccupied LLs, respectively. The blue, green, and red colors correspond to the first, second, and third groups of LLs.}
\end{figure}

\begin{figure}[htb]
\centering
\includegraphics[width=0.9\linewidth]{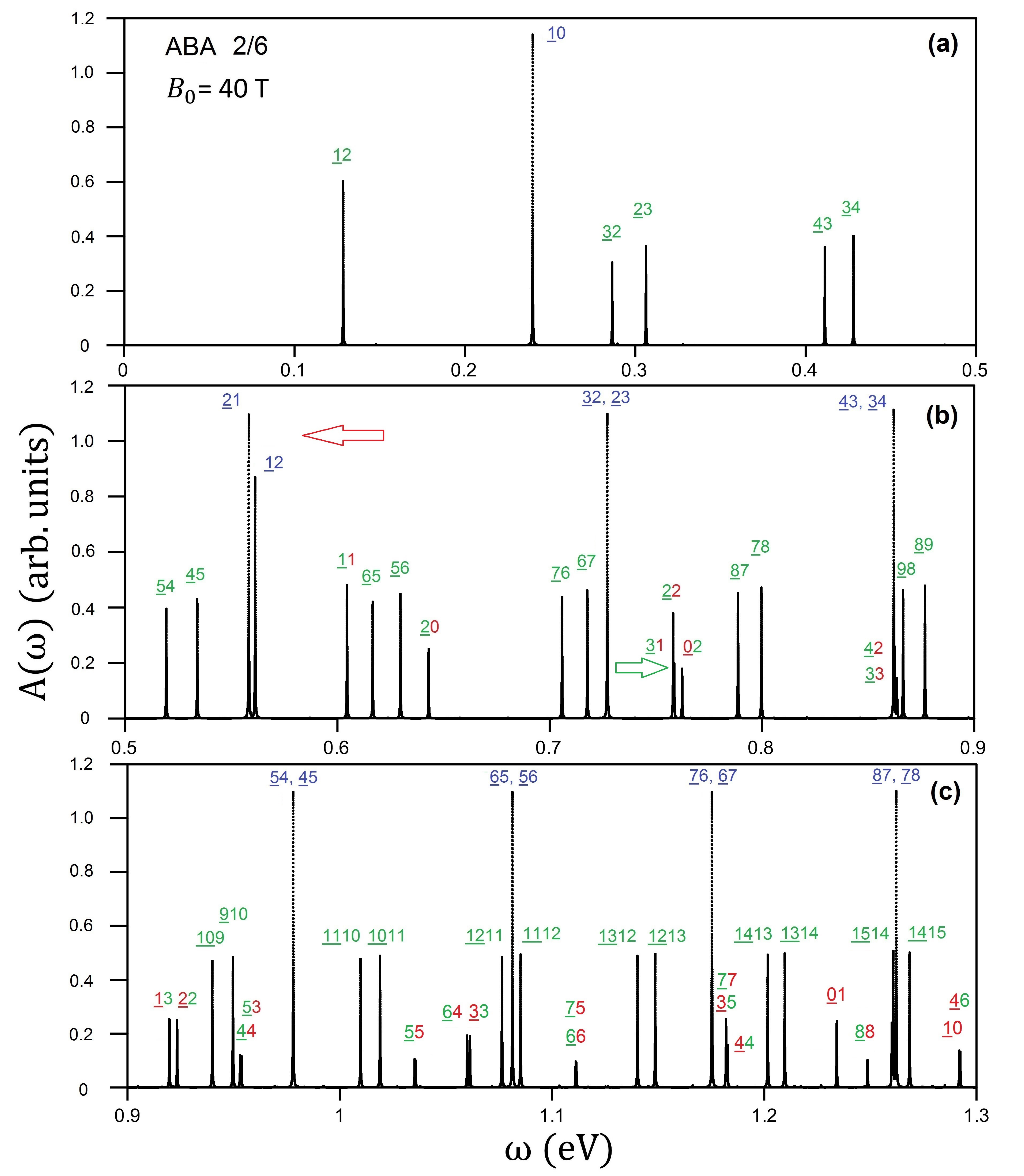}
\caption{Magneto-absorption spectra of AAB-stacked trilayer graphene in (a), (b), and (c) near 2/6 center with respect to different frequency ranges at $B_0=40$ T. The numbers with and without the underline stand for the quantum numbers of the occupied and unoccupied LLs, respectively. The blue, green, and red colors correspond to the first, second, and third groups of LLs. }
\end{figure}

\end{document}